# Distortion free pulse delay system using a pair of tunable white light cavities

H.N. Yum[1,3], M.E. Kim[2], Y.J. Jang[1] and M.S. Shahriar[1,2,*]

[1]*Department of Electrical Engineering and Computer Science, Northwestern University, Evanston, IL 60208, USA*
[2]*Department of Physics and Astronomy, Northwestern University, Evanston, IL 60208, USA*
[3]*Department of Electrical Engineering, Texas A&M University, College Station, TX 77843, USA*
[*]*Corresponding author: shahriar@northwestern.edu*

**Abstract:** Recently, a tunable bandwidth white light cavity (WLC) was demonstrated by using an anomalously dispersive intra-cavity medium to adjust a cavity linewidth without reducing the cavity buildup factor [G.S. Pati et al., Phys. Rev. Lett. **99**, 133601 (2007)]. In this paper, we show theoretically how such a WLC can be used to realize a distortion-free delay system for a data pulse. The system consists of two WLCs placed in series. Once the pulse has passed through them, the fast-light media in both WLCs are deactivated, so that each of these now acts as a very high reflectivity mirror. The data pulse bounces around between these mirrors, undergoing negligible attenuation per pass. The trapped pulse can be released by activating the fast-light medium in either WLC. Numerical simulations show that such a system can far exceed the delay-bandwidth constraint encountered in a typical data buffer employing slow light. We also show that the pulse remains virtually undistorted during the process.

## References and links

In recent years, there have been major breakthroughs in achieving slow-light, which results from a very steep positive dispersion, in a number of materials. Initially, electromagnetically induced transparency (EIT) was used in atomic gases [1] and solid-state material [2] for this purpose. EIT has also been observed in a solid at room temperature [3], as well as in semiconductor bands at 10K [4]. More recently, mechanisms other than EIT have been used to observe slow-light in room temperature solids. These include population pulsations in ruby [5] and semiconductor waveguides [6], stimulated Brillouin and Raman gain in fibers [7-11], as well as various resonators and periodic structures such as gratings and photonic crystals [12]. Some progress has also been made in enhancing the delay-bandwidth product (DBP) [13,14], as well as in the precise tuning of delays for individual bits [15]. So far, using slow-light techniques, a light pulse has been delayed by less than or several times the pulse duration [7,9,11]. Nonetheless, it has become clear of late that a slow-light based data buffering has severe limitations.

A more versatile technology that will allow data buffering with a high delay-bandwidth product is based on the use of dynamically reconfigured cavity bandwidth and dispersion in photonic crystals [16,17]. With high-speed electrical or optical control, the cavity finesse is lowered, thus enhancing its bandwidth to be large enough to load a high bit-rate data stream. Once the stream is loaded, the finesse is raised rapidly, thus trapping the data stream inside the cavity. However, physical implementation of this approach for practical systems has proved difficult for many reasons.

In this paper, we describe a variation of this approach by making use of the so-called white light cavity (WLC), which has a bandwidth much larger than the inverse of the cavity lifetime [18]. A pair of WLC's can be used to realize a trap-door type data buffer. Using Stimulated Brillouin Scattering (SBS) induced anomalous dispersion in an optical fiber, for example, such a buffer can be realized for wavelengths used in telecommunication. Unlike buffers that make use of slow light, the delay-bandwidth product for this WLC buffer can be very high, making it suitable for delaying very high bit rate data stream for a long time. Furthermore, the WLC buffer allows easy access to the data while inside the buffer, making it possible to perform on-line data sampling, processing and re-routing, using all optical techniques.

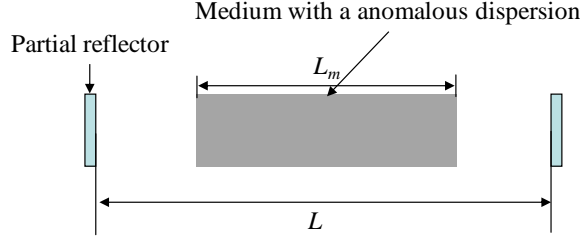

Fig.1. Schematic of a tunable-bandwidth WLC; Two partial reflectors, each with a reflectivity $R$, enclose the intracavity medium of length $L_m$ with anomalous dispersion, forming a cavity of physical length $L$.

Before presenting the details of this delay system, we briefly review a theoretical model to illustrate the relevant features of a WLC. Fig. 1 displays a typical WLC where two partial reflectors form a Fabry-Perot cavity of length $L$, containing an intracavity medium of length $L_m$ with anomalous dispersion. In what follows, we assume $L=L_m$, although this condition is not necessary. A monochromatic optical input wave, represented as $E_{in}(\omega) = E_0 e^{j\omega t}$, can be related to the output field as

$$E_{out} = E_0 e^{j\omega t} \times \frac{t^2 e^{-jkL}}{1 - r^2 e^{-2jkL}} \tag{1}$$

where $E_0$ is the field amplitude, $L$ is the cavity length, $t$ and $r$ are the field amplitude transmission and the reflection coefficients of each partial reflector, respectively, (for intensity, $R=r^2$, $T=t^2$ and $R+T=1$) and $\omega$ is the angular frequency of the optical field. Here, $k$ is the wave number, expressed as $k = n\omega/c$, where $n$ is the refractive index of the medium inside the cavity, and c is speed of light in vacuum. For the WLC, $n$ is a function of $\omega$, so that $k(\omega) = n(\omega)\omega/c$. We consider the particular case where the anomalous dispersion is produced by a gain doublet centered around $\omega_0$, the empty-cavity resonance frequency. This corresponds to an anti-symmetric profile for $n(\omega)$, so that $n_2 \equiv (1/2) d^2 n/d\omega^2 \big|_{\omega=\omega_0}$ has a null value. We can thus express $n(\omega)$ as a Taylor expansion around $\omega_0$: $n(\omega) = n_0 + (\omega - \omega_0) n_1 + (\omega - \omega_0)^3 n_3$, where $n_0$ is the index of the medium at $\omega = \omega_0$, $n_1 = dn/d\omega\big|_{\omega=\omega_0}$, and $n_3 = (1/6) d^3 n/d\omega^3 \big|_{\omega=\omega_0}$.

Under ideal conditions for a WLC, the product of the index and the frequency remains a constant: $n\omega = n_0 \omega_0$. If this condition holds for a range of $\Delta\omega$ around $\omega_0$ (assuming $\Delta\omega \ll \omega_0$), we have $n_1 \simeq -n_0/\omega_0$, and $n_3 \simeq 0$. Physically, it is easy to see what happens when the ideal WLC condition is fulfilled. The resonance condition for the cavity is $L = m\lambda/2$, where $m$ is an integer and $\lambda$ is wavelength, given by $\lambda = 2\pi c/(n\omega)$. Thus, if the product $n\omega$ remains a constant, the wavelength is independent of frequency, so that the cavity is resonant for all frequencies over the range $\Delta\omega$. A pulse with a bandwidth of $\Delta\omega$ will therefore transmit through the cavity resonantly, even though the empty cavity bandwidth may be much smaller than $\Delta\omega$. This is the essential concept behind the WLC. The consistency of the WLC effect with fast light is understood by considering the group index $n_g = n_0 + \omega_0 n_1$ and the group velocity $v_g = c/n_g$ of the intracavity medium. Under the ideal WLC condition, $n_1 = -n_0/\omega_0$ so that $n_g = 0$. This condition corresponds to an infinite group velocity of light; thus, the WLC effect is a manifestation of fast light.

More generally, the transfer function of WLC is given by $H_{WLC}(\omega) = t^2 e^{-jkL}/(1 - r^2 e^{-2jkL})$ where $k = \left[ n_0 + (\omega - \omega_0)n_1 + (\omega - \omega_0)^3 n_3 \right] \omega/c$. For the empty cavity, the transfer function can be written as $H_{EC}(\omega) = t^2 e^{-jk_0 L}/(1 - r^2 e^{-2jk_0 L})$ where $k_0 = \omega/c$. For completeness, we note that for free space propagation over a distance $L$, the transfer function in our notation is simply a phase accumulation: $H_{free}(\omega) = exp(-j\omega L/c)$.

Next, we consider the response to a pulse, $S_{in}(t)$. We choose the input pulse to be of the form $S_{in}(t) = exp(-t^2/t_0^2) exp\left[ j(\omega_0 + \xi)t \right]$, so that its Fourier transform is given by $\tilde{S}_{in}(\omega) = t_0/\sqrt{2} \, exp\left[ -\{(\omega - \omega_0 - \xi)t_0\}^2/4 \right]$. The carrier frequency of the pulse is upshifted by $\xi$ from the empty cavity resonance ($\omega_0$). By virtue of the convolution theorem, the output of a WLC is the inverse Fourier transform of the product of $\tilde{S}_{in}(\omega)$ and $H_{WLC}(\omega)$. Thus, the output intensity is $|S_{WLC}(t)|^2$, where $S_{WLC}(t) = 1/\sqrt{2\pi} \int_{-\infty}^{\infty} \tilde{S}_{in}(\omega) H_{WLC}(\omega) exp(j\omega t) d\omega$. Likewise, for a reference pulse that propagates in free space over the same distance, $L$, the resultant pulse after traveling is $S_{free}(t) = 1/\sqrt{2\pi} \int_{-\infty}^{\infty} \tilde{S}_{in}(\omega) H_{free}(\omega) exp(j\omega t) d\omega$.

Before computing the output signal, it is instructive to discuss the anticipated behavior by applying the well-known concept of group velocity to the whole system. The phase ($\phi$) for an individual frequency wave after a propagation distance of $L$ can be written as

$$\phi = \omega t - \frac{\omega n_{WLC} L}{c} \qquad (2)$$

where $n_{WLC}$ is the effective refractive index of the WLC. To find the group velocity (or index) of the WLC, we consider a pulse propagating through it. The pulse contains frequency components within a particular bandwidth, $\Delta\omega$. We require that after the propagation all the frequency components be added in phase [19] at the peak of the pulse, so that $d\phi/d\omega = 0$. We can then define the group velocity of the WLC simply as $v_{g(WLC)} = L/t$, where $t$ is the time of propagation through the length of the cavity under this constraint. After differentiating Eq. (2) with respect to $\omega$ and setting the result to equal zero, we find $v_{g(WLC)} = c/[d(n_{WLC}\omega)/d\omega]$. It is convenient to express the group velocity as $v_{g(WLC)} = c/n_{g(WLC)}$, where $n_{g(WLC)}$ is the effective group index of the WLC, which in turn is given by $n_{g(WLC)} = d(n_{WLC}\omega)/d\omega = n_{WLC} + \omega(dn_{WLC}/d\omega)$ [20].

If the complex transfer function $H_{WLC}$ is written as $|H_{WLC}| exp(j\angle H_{WLC})$, then the output of the WLC in frequency domain can be written as $\tilde{S}_{WLC}(\omega) = |H_{WLC}| exp(j\angle H_{WLC}) \tilde{S}_{in}(\omega)$. Here, $\angle H_{WLC}$ is the phase resulting from the propagation inside the WLC. Thus, the second term in Eq. (2) is given by $\omega n_{WLC} L/c = -\angle H_{WLC}$ so that $n_{WLC} = -c\angle H_{WLC}/(\omega L)$. We thus get:

$$n_{g(WLC)} = -\frac{c}{L}\frac{d\angle H_{WLC}}{d\omega} \qquad (3)$$

In general, the degree of pulse stretching in time domain after the propagation through the WLC is given by $\Delta T \approx (L/c)(dn_{g(WLC)}/d\omega)\Delta\omega$ where $\Delta\omega$ is the pulse bandwidth [19]. A positive (negative) value of $\Delta T$ corresponds to pulse broadening (compression). From Eq. (3), we obtain $\Delta T \approx -(d^2\angle H_{WLC}/d\omega^2)\Delta\omega$, so that the pulse maintains its original shape after propagation if its spectrum belongs to the spectral region where $d^2\angle H_{WLC}/d\omega^2 = 0$.

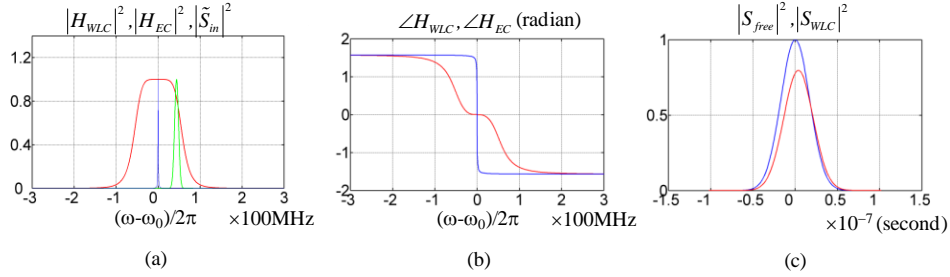

Fig. 2. (a) Transfer functions for empty cavity(blue) and for WLC(red), and the Fourier Transform of Gaussian input(green). $|\tilde{S}_{in}|^2$ is normalized to the peak magnitude of $|\tilde{S}_{in}|^2$. (b) Phases of $H_{EC}(\omega)$ (blue) and $H_{WLC}(\omega)$ (red). (c) $|S_{free}|^2$ (blue) and $|S_{WLC}|^2$ (red). The parameters of the intracavity medium are $n_1 = -8.223 \times 10^{-16}$/rad, $n_3 = 5.223 \times 10^{-35}$/rad$^3$.

Fig. 2 displays the transfer functions ($|H_{WLC}|^2, |H_{EC}|^2$), the phases ($\angle H_{WLC}, \angle H_{EC}$), the frequency spectrum of the input pulse ($|\tilde{S}_{in}|^2$) and the output pulses ($|S_{WLC}|^2$ and $|S_{free}|^2$), for a specific set of parameters. For the cavity, we have chosen the length $L=5$cm and the Finesse=999(R=0.999), so that the cavity bandwidth is about 2.9MHz (FWHM), and assumed, for simplicity of discussion, that the medium fills the whole cavity (i.e., $L_m = L$). We have used $\omega_0 = 2\pi \times 1.9355 \times 10^{14}$ corresponding to 1550nm which is a wavelength of widespread use in telecommunication. We have chosen the value of $n_1$ to satisfy the ideal WLC condition (i.e., $n_1 \simeq -n_0/\omega_0$). In a real system, such as the one employing dual gain peaks, it is easy to satisfy this condition. However, the value of $n_3$ is non-vanishing for such a system, thus limiting the bandwidth of the WLC. Here, we have chosen a value of $n_3$ corresponding to a WLC bandwidth of about 120 MHz (FWHM). The width of the Gaussian input pulse is chosen to be $\Delta \nu_{pulse}=29$MHz (corresponding to $t_0=34$ns) and the carrier frequency is shifted from $\omega_0$ by $\xi=1.5 \times \Delta \nu_{pulse}$. Thus, the frequency spectrum of the pulse is sufficiently separated from the spectral region of an empty cavity resonance. Explicitly, the pulse spectrum is expressed as $\tilde{S}_{in}(\omega) = t_0/\sqrt{2} \, exp\left[-\{(\omega-\omega_0-\xi)t_0\}^2/4\right]$. We will discuss later on the necessity for this shift in designing the trap-door data buffering system.

Fig. 2(a) indicates that for a WLC associated with $n_3=5.223 \times 10^{-35}$, the pulse spectrum $\tilde{S}_{in}(\omega)$ belongs to the spectral region where the amplitude of the cavity response function ($|H_{WLC}|$) begins to drop slightly from the uniform peak value. Thus, the intensity of the output pulse (red) is slightly smaller than its input value, as shown in Fig. 2(c). By comparing

$\left|\tilde{S}_{in}\right|^2$ to $\angle H_{WLC}$ in Fig. 2(b), we find that most of the spectral components of the probe is in the region where the slope is linearly approximated to $d\angle H_{WLC}/d\omega \simeq -2\times 10^{-8}$. Thus, according to Eq. (3), $n_{g(WLC)} \simeq 120$. As a consequence, the pulse slows down approximately $2\times 10^{-8}$ second compared to the reference which propagates in free space, as illustrated in Fig. 2(c). To understand the small pulse compression in time domain, we consider $d^2\angle H_{WLC}/d\omega^2$ and $\left|\tilde{S}_{in}\right|^2$. By inspecting Fig.2 (a) and (b), we find that $\left|\tilde{S}_{in}\right|^2$ contains the inflection point of $\angle H_{WLC}$ ( $d^2\angle H_{WLC}/d\omega^2 = 0$ ) which is close to the center of the pulse spectrum. The spectral components of the pulse are almost uniformly distributed within the spectral region of positive and negative values of $d^2\angle H_{WLC}/d\omega^2$, with a small shift towards the region where $d^2\angle H_{WLC}/d\omega^2 > 0$. Therefore, the output pulse is somewhat compressed in time, as can be seen in Fig. 2(c).

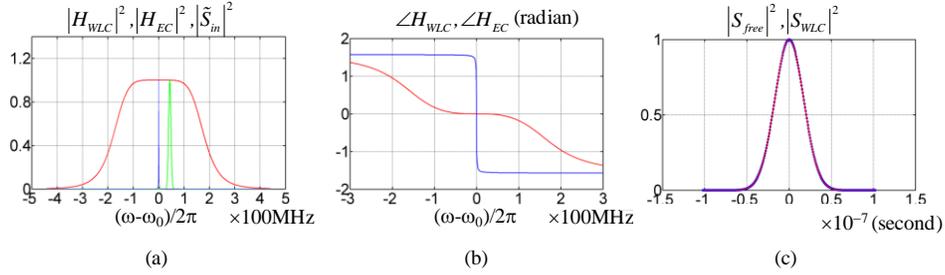

Fig. 3. For the medium with $n_1= -8.223\times 10^{-16}$/rad, $n_3=1.723\times 10^{-36}$/rad$^3$ (a) Transfer functions for empty cavity(blue) and for WLC(red), and the Fourier Transform of Gaussian input(green). $\left|\tilde{S}_{in}\right|^2$ is normalized to the peak magnitude of $\left|\tilde{S}_{in}\right|^2$. (b) Phases of transfer function for empty cavity(blue) and for WLC(red). (c) $|S_{free}|^2$(blue circles) and $|S_{WLC}|^2$ (red).

Fig. 3 illustrates the behavior of the WLC when the value of $n_3$ is reduced by a factor of 3, so that the bandwidth of the WLC is increased to nearly 350 MHz (see Fig. 3(a)). As indicated in Fig. 3(b), the input pulse spectrum now lies mostly within the region where $d\angle H_{WLC}/d\omega \simeq 0$ and $d^2\angle H_{WLC}/d\omega^2 = 0$. More specifically, $d\angle H_{WLC}/d\omega < 0$ so that $v_{g(WLC)} \gg c$, as dictated by Eq.(3). As a result, the pulse propagates through the WLC with virtually no reduction in amplitude and no distortion. Furthermore, it leads, by a very small duration, the pulse propagating in free-space. Given the relatively short length of the cavity, the degree of advancement is virtually unnoticeable here, as indicated by nearly overlapping pulses shown in Fig. 3(c). A more detailed analysis using numerical simulations, presented in Ref. [21], confirms this analytical result, and illustrates the behavior of the pulse as it passes through the WLC. Of course, as has been noted previously, such advancement does not violate causality [22]. Finally, note that in the absence of the dispersive medium, the pulse gets fully reflected by the cavity, since its spectrum is shifted from the empty cavity transmission window as shown in Fig.3(a).

From the results presented above, it is clear that a single WLC cannot be used to realize a data buffer. Instead, one must employ a pair of WLC's in series, in a trap-door configuration. Fig. 4 shows schematically the trap-door data buffering system, consisting of two identical WLCs. The one on the left is called the LWLC, and the one on the right is called the RWLC.

Here, each WLC is designed to be of the type described in Fig. 3 above. Each WLC has a length $L$, and has two partial reflectors (PRs) enclosing the dispersive medium inside. Each PR is assumed to be highly reflective, corresponding to a high finesse. The two WLCs are separated by a distance $L_2$. We assume that the negative dispersion inside each WLC is created by a pump signal with two frequency components (for example, in our original demonstration of a WLC [18], a Raman pump with two frequency components produced the double-peaked gain, yielding negative dispersion, with the gain negligibly small at the center). When the pump is turned off, the dispersion vanishes, and the WLC is converted to an ordinary cavity. The frequency separation of the two components inside the pump and the intensities thereof are parameters which determine the slope of the dispersion. One can control theses parameters to change $n_1$ and $n_3$ and thus manipulate the linewidth of each WLC.

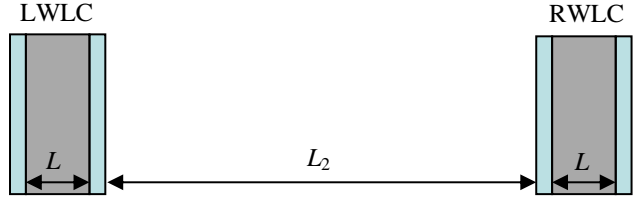

Fig. 4. Diagram of the proposed pulse delay system. Two identical WLCs are separated by a distance of $L_2$.

The data buffering process works as follows. Consider a Gaussian data pulse with a bandwidth ($\Delta \nu_{pulse}$) that is broader than the linewidth of the empty cavity for each WLC. We also assume the carrier frequency of the data pulse to be shifted by $1.5 \times \Delta \nu_{pulse}$ from the resonance frequency ($\omega_0$) of each empty cavity. Ordinarily, the WLC process is kept turned off in each WLC. Once the buffer is ready to load the pulse, entering from the left, the WLC effect is activated in LWLC only. As illustrated in $S_{cavity}(t) \& S_{free}(t)$ of Fig. 3(c), the pulse appears at the exit of the LWLC without any delay, attenuation or distortion compared to a reference pulse which propagates in free space. Once the pulse has left the LWLC, the WLC effect inside LWLC is turned off. We assume that the intermediate zone between LWLC and RWLC is long enough to spatially confine the data pulse. The pulse now propagates in the middle of the two WLCs and reaches the left PR of RWLC. Note that since the WLC effect in the RWLC is kept turned off, it behaves as an empty cavity. Since the pulse is sufficiently shifted from the resonance frequency $\omega_0$ of the empty cavity, the frequency spectrum of the pulse has a negligible overlap with the narrow transmission spectrum ($\Delta \nu_{FWHM} = 2.9 MHz$) of the empty cavity, as illustrated in Fig. 3(a). As such, the RWLC now acts as a simple reflector with the very high reflectivity ($R$) of the PR. The pulse is therefore reflected with only a small attenuation due to the finite transmission of the PR. When the pulse reaches the LWLC after reflection, the LWLC also acts like a near perfect reflector with a high reflectivity of R, since the WLC effect inside it has been turned off.

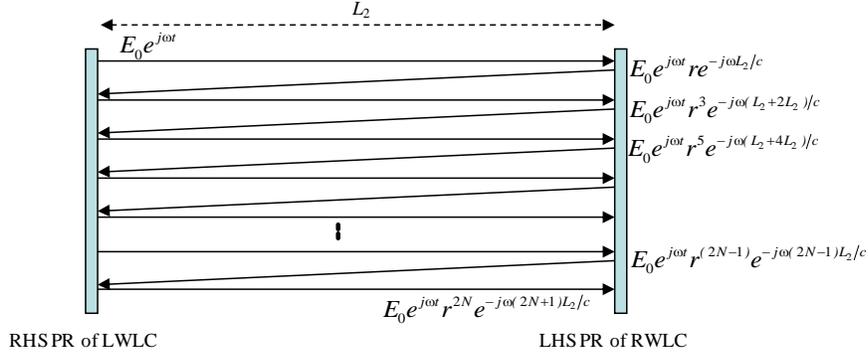

Fig. 5: Illustration of N round trips between two reflectors. See text for details.

The pulse will thus remain trapped between the two WLCs, bouncing between them. Once we are ready to release the pulse, we activate the WLC effect in the RWLC, for example. Upon reaching the RWLC, the pulse will now pass through it, again with virtually no attenuation or distortion. The net delay achieved is simply given by the number of bounces times the round trip time between the two WLCs. The limit on the maximum number of bounces is determined by the residual attenuation due to the very small but finite transmission of the PRs.

The discussion presented above can be summarized by finding a transfer function for the entire system which is composed of the LWLC, the RWLC and the region in the middle. First, consider the transfer function of the intermediate zone. As illustrated in Fig. 5, consider a monochromatic wave starting from the right PR of LWLC (PR1) and arrives at the left PR of RWLC (PR2) after N round trips. If $E_0 e^{j\omega t}$ represents the wave at PR1, the wave at PR2 after N round-trips can be is written as $E_0 e^{j\omega t} R^N e^{-j\omega(2N+1)L_2/c}$ where $L_2$ is the distance between the two PRs. Therefore, the transfer function of the intermediate zone can be written as:

$$H_i(\omega) = R^N e^{-j\omega(2N+1)L_2/c} \qquad (4)$$

The total transfer function $H_{total}(\omega)$ of the pulse delay system can be expressed as $H_{total}(\omega) = H_L(\omega) H_i(\omega) H_R(\omega)$ where $H_R(\omega)$ and $H_L(\omega)$ are the transfer functions of the RWLC and the LWLC, respectively. Since the WLCs are identical, we have $H_R(\omega) = H_L(\omega) = H_{WLC}(\omega)$. Again, by virtue of the convolution theorem, the output of the delay system can be written as

$$S_{system}(t) = \frac{1}{\sqrt{2\pi}} \int_{-\infty}^{\infty} H_{total}(\omega) \tilde{S}_{in}(\omega) exp(j\omega t) d\omega \qquad (5)$$

In order to determine the net delay, it is also necessary to compute the transfer function of the reference pulse, which propagates in free space over the distance of $(2L+L_2)$. As before, this propagation can be represented by the simple transfer function of $H_{free}(\omega) = exp[-j\omega(2L+L_2)/c]$, so that $S_{free}(t) = 1/\sqrt{2\pi} \int_{-\infty}^{\infty} H_{free}(\omega) \tilde{S}_{in}(\omega) exp(j\omega t) d\omega$.

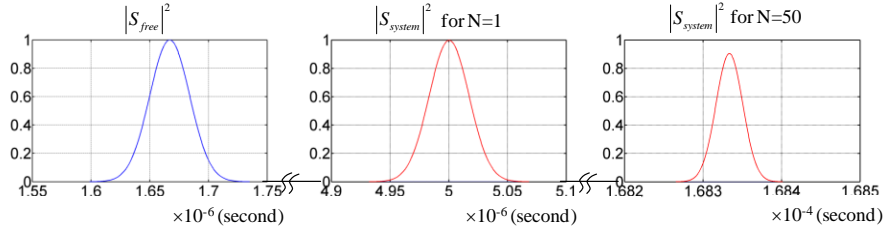

Fig. 6. At *t*=0, the reference and the data pulses are launched at the entrance of the LWLC. Blue is the reference pulse ( $S_{free}(t)$ ). It propagates the optical path of $2L+L_2$ in free space and the center of the pulse appears at the exit of RWLC after $t=(2L+L_2)/c \approx 1.67\times10^{-6}\,second$. The data pulse is observed at the output of the RWLC after $t=(2L+3L_2)/c \approx 5.00\times10^{-6}\,second$ for one round trip (N=1) and $t=(2L+101\times L_2)/c \approx 1.6833\times10^{-4}\,second$ for N=50.

Fig. 6 graphically illustrates $S_{free}(t)$ and $S_{system}(t)$. Recall that we have chosen the input pulse to be the same as the Gaussian pulse in Fig. 3(a) and are using the WLC parameters in Fig. 3(b) for both the RWLC and the LWLC. Specifically, we have used a data pulse of $\Delta\nu_{pulse} = 29\,MHz$. The carrier frequency of the pulse is shifted from the empty cavity resonance by $1.5\times\Delta\nu_{pulse} = 33.5\,MHz$. For illustration, we consider $L_2$=500m. The numerical simulation suggests that for one round trip (N=1), the data pulse is delayed by $2L_2/c = 3.3\times10^{-6}\,sec$. For fifty round trips (N=50), the delay time is observed to be $100L_2/c = 1.67\times10^{-4}\,sec$. Note that the delay is equal to approximately 4843 times the input pulse duration ($1.7\times10^{-5}/t_0 \simeq 4843$). The intensity attenuation is less than 10% due to the finite transmission of the PRs. Of course, the attenuation can be further reduced by using PRs with even higher reflectivities. The large value of $L_2$ can be realized by an optical fiber loop, for example. This loop can be coupled to the WLCs with negligible loss if the WLCs are realized using fast light effect in optical fiber as well [23]. Of course, there would be an additional attenuation of 0.2dB/km for 1550nm. Taking into account the typical refractive index of about 1.46, this would correspond to an additional attenuation of ~6.8dB. This is actually less than the attenuation suffered by optical data in long haul transmission, where the typical distance between regenerators could be as high as 100 km. Since the attenuation is not expected to be accompanied by any significant distortion, it should be easy to restore the pulse amplitude by using an optical amplifier, for example. In Ref. 23, we show an explicit scheme for employing an amplifier for restoring the pulse amplitude in conjunction with the WLC data buffer, and point out the significant advantages of this approach over the conventional recirculating buffer [24].

The bandwidth of the delay system can be defined as the maximum frequency spectrum width of a data pulse that the system can delay without noticeable distortion. In this case, the effective bandwidth is given by half of the difference between the WLC bandwidth and the empty cavity bandwidth. For the parameters used in Fig. 3, the WLC bandwidth is about 350 MHz and the empty cavity bandwidth is about 2.9 MHz, corresponding to a delay system bandwidth of about 173.5 MHz. The largest bandwidth achievable for the WLC depends on the mechanism used for producing the negative dispersion. We have shown that a bandwidth of ~21 GHz may be possible using double-peaked Brillouin gain in optical fiber under particular conditions [23].

A very important feature of this design is that the delay time is independent of the system bandwidth, and is determined simply by the time elapsed inside the region between the LWLC and the RWLC. For the result shown in Fig. 6, the delay-bandwidth product is nearly

$2.9 \times 10^4$, which is well beyond the typical values achievable using conventional delay systems. As mentioned above, the delay time can be increased even more by using PRs with higher reflectivities, or by allowing for higher degree of distortion-free attenuation, thereby yielding a much higher value of the delay-bandwidth product.

In conclusion, we show theoretically how a pair of WLCs can be used to realize a distortion-free delay system for a data pulse. Numerical simulations show that such a system can far exceed the delay-bandwidth constraint encountered in a typical data buffer employing slow light. We also show that the pulse remains virtually undistorted during the process.

This work was supported by DARPA through the slow light program under grant FA9550-07-C-0030, by AFOSR under grants FA9550-06-1-0466 and FA9550-10-1-0228, and by the NSF IGERT program under grant DGE-0801685.